\documentclass[prl,aps,twocolumn,showpacs]{revtex4}
\usepackage{graphicx,epsfig}

\def\be{\begin{equation}}
\def\ee{\end{equation}}
\def\bea{\begin{eqnarray}}
\def\eea{\end{eqnarray}}

\begin{document}

\title{Maximum entropy and the problem of moments: A stable algorithm} 

\author{K.\,Bandyopadhyay and A.\,K.\,Bhattacharya}
\email{physakb@yahoo.com}
\affiliation{Department of Physics, University of Burdwan, Burdwan, WB 713104, India}

\author{Parthapratim Biswas} 
\email{biswas@phy.ohiou.edu}
\author{D.\,A.\,Drabold}
\email{drabold@ohio.edu}
\affiliation{Department of Physics and Astronomy, Ohio University, Athens, OH 45701}


\begin{abstract}

We present a technique for entropy optimization to calculate a distribution from 
its moments. The technique is based upon maximizing a discretized form of the 
Shannon entropy functional by mapping the problem onto a dual space where an 
optimal solution can be constructed iteratively. We demonstrate the performance 
and stability of our algorithm with several tests on numerically difficult functions. 
We then consider an electronic structure application, the electronic density of 
states of amorphous silica and study the convergence of Fermi level with increasing 
number of moments. 
\end{abstract} 

\maketitle

One of the fixed themes of physics is the solution of inverse problems.
A ubiquitous example in theoretical physics is the ``Classical Moment
Problem" (CMP), in which only a finite set of power moments of a
non-negative distribution function $p$ is known, and the full
distribution is needed~\cite{Shohat}. It is obvious that the solution for $p$ is {\it
not unique} for a finite set of moments. This non-uniqueness suggests
the need for a ``best guess" for $p$, based upon the available
information.  With its ultimate roots in nineteenth century statistical
mechanics and a subsequent strong justification based upon probability
theory, the ``maximum entropy" (maxent) method has provided an
extremely successful variational principle to address this type of inverse 
problem~\cite{Jaynes}. Collins and Wragg used the maxent method to solve 
the CMP for a modest number of moments~\cite{Collins}. In a 
comprehensive paper with seminal applications, Mead and Papanicolaou~\cite{Mead1} 
solved the CMP with
maximum entropy techniques and proposed the first practical numerical
scheme to solve the moment problem for up to 15 moments. In a
host of subsequent papers, the utility of the method as an unbiased and
surprisingly efficient (rapidly convergent) solution of the CMP has
been established. The principle has been used extensively in a number of 
diverse applications ranging from image construction to spectral analysis, 
large-scale electronic structure problems~\cite{Drabold1,Silver0}, series 
extrapolation and analytic continuation~\cite{Drabold2}, quantum electronic transport~\cite{Mello}, 
ligand-binding distribution in polymers~\cite{Poland}, and transport 
planning~\cite{Steeb}.

There exist a number of maximum entropy algorithms~\cite{Skilling, Mead1, Turek, Silver0, Brett} 
that have been developed over the last two decades. Many of the algorithms 
(but not all) are constrained  by the number of moments that it can deal with 
and become unreliable when the number of constraints exceeds a problem-dependent 
upper limit. As the number of moments increases, the calculation of moments 
(particularly the power moments) becomes more sensitive to machine 
precision and the optimization problem becomes ill-conditioned. It has 
been observed that implementation of a maxent  algorithm with more than 20 
power moments is notoriously difficult even with extended precision arithmetic 
and it rarely gives any further information on the nature of the distribution. 
The use of orthogonal polynomials as basis set significantly improves the 
accuracy and remedies most of the problems that one encounters with power 
moments. 

In this paper we present an iterative approach to construct the maxent 
solution of CMP, which is based upon discretization of the Shannon entropy 
functional~\cite{Shannon}. 
The essential idea is to discretize Shannon entropy
and map the problem from the primal space onto dual space where an 
optimal solution can be constructed iteratively without 
the need of matrix inversion. We discuss theoretical ideas and develop 
algorithms that can be used with both power and Chebyshev moments. The 
stability and the accuracy of the method are discussed with reference to 
two numerically non-trivial examples -- a uniform distribution and a double-delta function. 
We illustrate the usefulness of our technique by computing the 
electronic density of states (EDOS) of amorphous silica with 
particular emphasis on convergence of the Fermi level as a function of number 
of moments.

The starting point of our approach is to use a discretized 
form of the Shannon entropy functional~\cite{Shannon} $S[x]$
using a quadrature formula 
\be 
\label{eq-010}
S = - \int \, p(x) \ln p(x) dx \approx - \sum_{j=1}^n w_j \, p_j \ln p_j 
\ee

Here $ w_j$  and $ x_j$ are the weights and abscissas of any accurate 
quadrature formula, say the Gauss-Legendre and without any loss of 
generality we restrict ourselves to $x \in $ [0,1].  We want to maximize 
$S$ subject to the discretized moment constraints

\be 
\label{eq-020}
\sum_{j=1}^{n} w_j\, x_j^i \, p_j = \sum_{j=1}^n a_{ij} \tilde p_j = \mu_i, \; i = 1, 2, ..., m 
\ee 

where we define $\tilde p_j = w_j\, p_j $ and $ a_{ij} = x_j^i $. 
The entropy optimization program (EOP) can now be stated as to optimize 
the Lagrangian function 
\be 
\label{eq-050}
L({\bf \tilde p}, \eta) \equiv \sum_{j=1}^n \tilde p_j \, \ln \left(\frac{\tilde p_j}{w_j}\right)
- \sum_{i=1}^m \tilde \eta_i \left(\sum_{j=1}^n a_{ij} \tilde p_j - \mu_i \right) 
\ee

and the solution can be written as

\be 
\label{eq-060}
\tilde p_j = w_j \exp\left(\sum_{i=1}^m a_{ij} \tilde \eta_i - 1 \right), \: \: j = 1, 2,..., n
\ee 

Since  ${\bf w} \ge 0$, Eq.(\ref{eq-060}) implies that  ${\bf \tilde p} \ge $ 0. 
Furthermore, the conditions in Eqs.(\ref{eq-020}) and (\ref{eq-060}) can be 
lumped together

\be 
\label{eq-070}
h_i(\tilde \eta) \equiv \sum_{j=1}^n a_{ij} \, w_j \, \exp \left(\sum_{k=1}^m a_{kj} 
\tilde \eta_k - 1 \right) - \mu_i = 0,  \; \; \forall \;  i. 
\ee 

We now see from Eq.(\ref{eq-070}) that the original constrained optimization 
program is now reduced to an {\em unconstrained convex optimization program} 
involving the dual variables

\be
\label{eq-080}
\min_{\tilde \eta \in R^m} \: d(\tilde \eta) \equiv  \sum_{j=1}^n  w_j \exp 
(\sum_{i=1}^m a_{ij}\tilde \eta_{i} - 1) - \sum_{i=1}^m \mu_i \tilde \eta_i
\ee 

If the dual optimization program stated above has an optimal solution 
${\tilde {\bf \eta^*}} $, the solution ${\tilde p_j ({\bf \eta^*})}$ 
can be obtained from Eq.(\ref{eq-060}). Bergman has proposed an iterative 
method to minimize the dual objective function $d(\bf \tilde \eta)$ 
taking {\em only one} dual variable at a time~\cite{Bergman}. The method 
starts with an arbitrarily chosen ${\bf \tilde \eta^0} \in R^m$, and 
then cyclically updates all the dual variables as follows: 

Step 1: Start with any ${\bf \tilde \eta^0} \in R^m$  and a sufficiently small 
tolerance level $\epsilon >$ 0. Set k = 0  and obtain $\tilde p_j^0$. 

Step 2: Let i =  (k mod m) + 1. Solve the equation
 
\bea 
\label{eq-105}
\phi_i^k(\lambda^k)   = \sum_{j=1}^n a_{ij} \tilde p_j^k \exp(a_{ij}\lambda^k) - \mu_i = 0 
\eea 

Step 3: Update each component of ${\bf \tilde \eta}$ 

\be 
\label{eq-110}
\tilde \eta_l^{k+1}  = \tilde \eta_l^k + \lambda^k (\mbox{if}\; l = i),  \; \; 
\tilde \eta_l^{k+1} = \tilde \eta_l^k \; \mbox{if} \; l \ne i 
\ee 

Step 4: If Eq.(\ref{eq-070}) is satisfied within the preset level of 
tolerance, then stop with ${\bf \eta^*} = {\tilde \eta^k}$,  
and obtain the primal solution from Eq.(\ref{eq-060}). Otherwise, 
calculate 
\be 
\tilde p_j^{k+1} = w_j \exp(\sum_{i=1}^m a_{ij} \tilde 
\eta_i^{k+1} -1), \; \; \; j = 1,2,...,n
\ee
and go to Step 2 

From a computational point of view, the most problematic part of the above 
algorithm  is the solution of the set of Eq.(\ref{eq-105}) in Step 2. 
In a variant of the above scheme known as multiplicative algebraic reconstruction 
technique~\cite{Fang, Gordon}, one uses the following closed-form expression 
to approximate the correction term $\lambda^k$
\be
\label{eq-120}
\lambda_i^k = \ln \left(\frac{\mu_i}{\sum_{j=1}^n a_{ij}\tilde p_j^k}\right)
\ee

Step 3 of the algorithm is now modified by substituting the expression 
above for $\lambda_i^k$ in Eq.(\ref{eq-110}). A convergence theorem for the 
modified algorithm can be found in Lent~\cite{Lent}. It is, however, 
quite easy to see that the algorithm will fail unless for every 
$i = 1, 2, ., m,$  either

\be
\label{eq-130} 
\mu_i > 0 \;\;\; \; \mbox{and} \;\;\; 0 \le a_{ij} \le 1, \; \; j = 1, 2, ..., n 
\ee
{\hskip 4cm or} 
\be
\label{eq-140} 
\mu_i < 0 \;\;\; \; \mbox{and} \;\;\; 0 \ge a_{ij} \ge -1, \; \; j = 1, 2, ..., n
\ee 

We note that in this case we are assured of convergence of the solution 
of our discretized EOP because the condition (\ref{eq-130}) holds.

The EOP algorithm above can only be used provided that the condition 
stated by the inequality (\ref{eq-130}) or (\ref{eq-140}) is satisfied. 
This constrains us to apply the algorithm for power moments but 
neither of these two are necessarily true for other polynomial moments. 
In order to work with Chebyshev polynomials, we first employ the averages 
of shifted Chebyshev polynomials~\cite{num-recipe} of the first kind 
$T_n^{*}(x) = T_n(2x-1)$ to recast the entropy optimization program 
(EOP) given by statement (\ref{eq-070}). The only change needed for 
this purpose is to redefine $a_{ij}$ by $a_{ij} = T_i^{*}(x_j)$.  

Our next step is to find a transformation that will convert 
the EOP into an equivalent problem in which all the program 
parameters are non-negative. For finding the necessary transformation, we 
define for $i = 1, 2, 3, . . ., m,$ 
\be
u_j = [\max_j (-a_{ij}) ] + 1. \nonumber 
\ee 
Obviously, for $i = 1, 2, 3, . . . , m$ and $j = 1, 2, 3, . . . , n,$ 
\be
(u_i + a_{ij}) > 0.  \nonumber 
\ee 

Let us now define for $i = 1, 2, 3, . . . , m,$
\be 
M_i \equiv \max_j(u_j + a_{ij}) \: \: ; \: \: t_i \equiv \frac{1}{m(M_i+1)}. \nonumber 
\ee

It is easy to see that the following relations hold for $i = 1, 2, 3, . . . , m$

\bea
&&M_i > 0, \; \; t_j  > 0 \nonumber \\ 
&&(M_i+1)\,t_j = \frac{1}{m}, \; \; t_i\,(u_i + a_{ij}) \le t_i\,M_i < \frac{1}{m} 
\nonumber 
\eea

For $i = 1, 2, 3, . . . , m$ and $j = 1, 2, 3, . . . , n,$ let us define 
\be 
a_{ij}^{'} \equiv t_i(u_i + a_{ij}). 
\ee
Apparently, for $i = 1, 2, 3, . . . , m$ and $j = 1, 2, 3, . . . , n,$ we have 
\be 
\frac{1}{m} > a_{ij}^{'} > 0 \: \: ; \: \: 0 < \sum_{i=1}^m a_{ij}^{'} = 
\sum_{i=1}^m t_i(u_i + a_{ij}) < 1. 
\ee

It is interesting to note that if $ {\bf \tilde p}$ is a feasible solution to 
the EOP involving averages of $T_n^{*}(x)$, then for $i = 1, 2, 3, . . . , m$
\be 
\sum_{j=1}^n a_{ij}^{'}\tilde p_j = \sum_{j=1}^n t_i(u_i + a_{ij}) \tilde p_j = t_i(u_i 
+ \mu_i). 
\ee 
Hence, if we define for $i = 1, 2, 3, . . . , m,$
\be 
\label{eq-new}
\mu_i^{'} \equiv t_i\,(u_i + \mu_i) = \sum_{j=1}^n a_{ij}^{'} \, \tilde p_j.
\ee 

It is easy to verify that $ 1/m > \mu_i^{'}$ for $ i = 1, 2, 3, . . ., m.$ 
The transformed EOP has thus the same form as previously, except for the fact 
that we use Eq.(\ref{eq-new}) in place of Eq.(\ref{eq-020}). Since both 
$a_{ij}^{'}$ and $ \mu_i^{'}$ can take only positive values, a feasible 
solution to the original program  can now be obtained by replacing $a_{ij}$ 
and $\mu_{ij}$ in Eq.(\ref{eq-070}) by $a_{ij}^{'}$ and 
$\mu_{ij}^{'}$~\cite{note1}. 

We consider two numerically difficult examples, a uniform distribution and a 
double-delta function, to study the stability and accuracy of the algorithm.  
The Chebyshev moments of these two functions can be exactly calculated. Earlier 
efforts to reproduce these distributions have met with limited success because 
of the difficulty in matching a sufficient number of moments and for the singular
nature of the functions. It would be interesting to see how the algorithm performs 
in case of a) Uniform distribution $f(x) = 1 $,  $x \in [0,1]$ and b) a 
double-delta function $g(x) = \delta (x-\frac{1}{4})+\delta(x-\frac{3}{4})$, $ x \in [0,1]$. 

\begin{figure}
\includegraphics[width=2.25in,height=2.25in,angle=270]{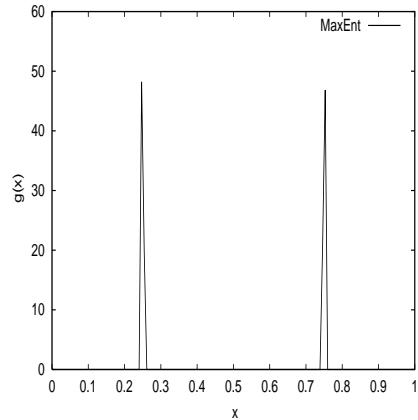}
\caption{
\label{fig1}
Reconstruction of a double-delta function $ {\rm g(x)=\delta(x-\frac{1}{4})
+\delta(x-\frac{3}{4}})$ from shifted Chebyshev moments. 
}
\end{figure}

The algorithm produces the uniform distribution correctly up to five decimal 
places. We found that the first 25 shifted Chebyshev moments are sufficient 
for this purpose. The fact that the end points have been produced so accurately 
without any spurious oscillations is a definitive strength of this approach 
and reflects the stability and accuracy of our algorithm. In figure \ref{fig1}, 
we have plotted 
the result for the double-delta function.  The result is equally convincing and 
certainly establishes the usefulness of this method over the other existing ones 
in the literature. 
In addition to these examples, we have also tested our algorithm to reconstruct a 
Tent map, a semicircular distribution, a square-root distribution and a distribution 
with a gap in the spectrum. In all these cases, the algorithm correctly produces 
all the features of the distributions without failing. These results clearly 
demonstrate that the algorithm is very stable, accurate and is capable 
of producing some very uncommon distributions (such as double-delta function) without 
any difficulty. 

We now consider a practical case where exact moments are not known but approximate 
moments are available. An archetypal example is the calculation of electronic density 
of states from its moments.  In the context of solid state physics, maxent 
has been used profitably to calculate the density of electronic (vibrational) states 
from a knowledge of the moments of the Hamiltonian (Dynamical) matrix. The computation 
of moments itself is an interesting problem in this field and there are methods 
available in the literature that specifically address this 
issue~\cite{Drabold1, Skilling}. 
\begin{figure}
\includegraphics[width=2.25in,height=2.25in,angle=270]{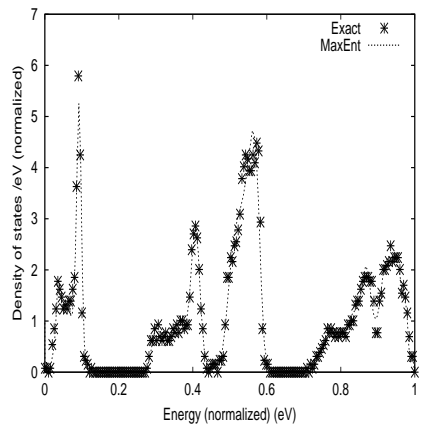}
\caption{
\label{fig2}
Normalized electronic density of states/eV (dotted line) of amorphous silica 
using the first 60 shifted Chebyshev moments. The distribution of energy 
eigenvalues (point) from direct diagonalization of the Hamiltonian 
matrix is also plotted in the figure. Normalized Fermi level is at 
0.595 eV. 
}
\end{figure}
Here one is interested in determining physical quantities such as Fermi level 
and band energy of large systems (e.g.~clusters, biological macromolecules etc.) 
without diagonalizing the Hamiltonian matrix. For amorphous semiconductors, 
this is particularly suitable because of disordered scattering (of electrons) that 
washes out the van Hove singularities in the electronic spectrum. A stable and 
accurate maxent algorithm, therefore, would be very useful in calculating 
electronic properties of amorphous semiconductors. The two examples discussed 
above suggest that we should be able to produce complex electronic spectrum with 
a gap (or gaps) to a high degree of precision and hence the Fermi level and band 
energy. As for metallic systems, the determination of Fermi energy is a non-trivial 
problem for $O(n)$ methods. A primary requirement for a maxent algorithm in this 
case is that 1) it must produce the distribution accurately and 2) it must
do so in a stable way using a sufficient number of moments to correctly produce 
the singularities of the spectrum. It is very pleasing to note that our algorithm 
does satisfy this requirement and therefore may offer an alternative approach to 
compute Fermi energy of metallic systems. 

\begin{figure}
\includegraphics[width=2.25in,height=2.25in,angle=270]{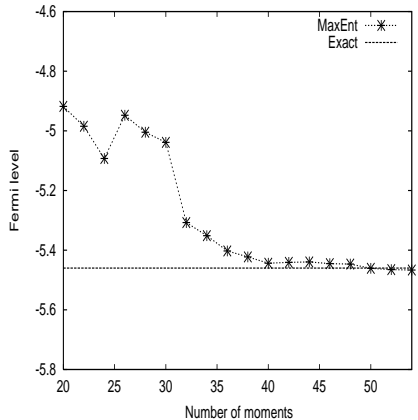}
\caption{
\label{fig3}
Fermi level of amorphous silica as a function of number of the shifted 
Chebyshev moments.  The value obtained from direct diagonalization of 
the Hamiltonian matrix is -5.465 eV and is plotted as a horizontal 
line in the figure. 
}
\end{figure}

In figure~\ref{fig2}, we have plotted the EDOS of amorphous silica using first 60 
moments and compared it to the result obtained by direct diagonalization of the 
Hamiltonian matrix. It is clear from the figure that all the features of the EDOS 
are correctly produced by our maxent algorithm. Finally, in figure~\ref{fig3} we 
have plotted the variation of Fermi energy with the number of moments.  The Fermi 
energy is computed by integrating the normalized density of states to obtain 
the correct number of total electrons. It is clear from figure~\ref{fig3} that 
the Fermi energy starts to converge after first 30 moments and eventually converges 
after 40 moments. 

In conclusion, we present an algorithm for maximum entropy construction of a 
distribution from its moments. The algorithm is very stable, accurate and can 
handle a large number of moments~\cite{note2} (up to 500). The usefulness of this algorithm 
is demonstrated by constructing some numerically difficult distributions and 
applying it to amorphous silica to compute the electronic density of states 
and the Fermi level. 

We acknowledge the support of National Science Foundation under Grant 
Nos.\,DMR-0205858 and DMR-0310933.


\begin{thebibliography}{*99}

\bibitem{Shohat}
J. A. Shohat and J. D. Tamarkin, {\em The Problem of Moments}, 
(American Mathematical Society, Providence, Rhode Island, 1963). 

\bibitem{Jaynes}
E.T. Jaynes, {\em Probability Theory: The Logic of Science} (Cambridge University Press, 2003) 

\bibitem{Collins}
R. Collins and A. Wragg, J Phys. A: Math. Gen. 10, 1441 (1977)

\bibitem{Mead1}
L. R. Mead and N. Papanicolaou, J. Math. Phys. {\bf 25}, 2404 (1984). 

\bibitem{Drabold1}
D. A. Drabold and O. F. Sankey, Phys. Rev. Lett. 70, 3631 (1993). 

\bibitem{Silver0}
R.N. Silver and H. R\"oder,  Phys. Rev. E 56, 4822 (1997)

\bibitem{Drabold2}
D.A. Drabold and G.L. Jones, J. Phys. A: Math. Gen. 24, 4705 (1991) 

\bibitem{Mello}
P.A. Mello and Jean-Louis Picard, Phys. Rev. B {\bf 40}, R5276 (1989)

\bibitem{Poland}
D. Poland, J. Chem. Phys. 113, 4774 (2000)

\bibitem{Steeb}
W-H Steeb, F. Solms and R.Stoop, J. Phys. A: Math. Gen. 27, L399 (1994). 

\bibitem{Skilling}
J. Skilling, in {\em Maximum entropy and Bayesian Methods}, edited 
by J. Skilling (Kluwer, Dordrecht, 1989)

\bibitem{Turek}
I. Turek, J. Phys. C: Solid St. Phys. 21, 3251 (1988). 

\bibitem{Brett} 
G. L. Bretthorst (Unpublished)

\bibitem{Shannon}
C.\,Shannon, Bell System Tech J. {\bf 27}, 379 (1948)

\bibitem{Bergman}
L. M. Bergman, U.S.S.R. Comput. Maths. and Math. Phys. 7, 200 (1967). 

\bibitem{Fang}
S.C.Fang, J.R.Rajasekara, and H. -S. J. Tsao, {\em Entropy Optimization 
and Mathematical programming}, (Kluwer Academic Publishers, Dordrecht, 1997). 

\bibitem{Gordon}
R. Gordon, R. Bender and G. T. Herman, J. Theoret. Biol. 29, 471 (1970). 

\bibitem{Lent}
A. Lent, in {\em Image analysis and evaluation}, edited by R. Shaw (SPSE, 
Washington, D. C. 1953). 

\bibitem{num-recipe}
M. Abramowitz and I. A. Stegun, {\em Handbook of mathematical 
functions}, (Dover Publications, New York, 1972). 

\bibitem{note1}
The transformed problem in terms of $a_{ij}^{'}$ and $\mu_i^{'}$ 
has exactly the same solution as the original problem. If the original 
problem is infeasible (due to inaccurate values of higher power moments 
etc.), this gets reflected by the lack of positive definiteness of 
$a_{ij}^{'}$ and $\mu_i^{'}$. 

\bibitem{note2}
In principle there is no limit to the number of moments that can be handled 
by the method at the expense of computational time. In the present context we 
have gone up to 500 moments without any difficulty. 

\end{thebibliography}
\end{document}